\documentstyle[preprint,aps,epsf]{revtex}

\begin{document}
\draft
\title{
A Momentum Transformation Connecting a NN Potential in the Nonrelativistic 
and the Relativistic Two-Nucleon Schr\"odinger Equation
}

\author{H. Kamada\footnote{present address: Institut f\"ur
    Kernphysik, Fachbereich 5 der Technischen Hochschule Darmstadt, D-64289
    Darmstadt, Germany   } and W. Gl\"ockle}
\address{Institut f\"ur Theoretische Physik II, 
Ruhr-Universit\"at Bochum, D-44780 Bochum, Germany}

\date{\today}
\maketitle

\begin{abstract}
An analytical  relation between  center of mass  momenta in a nonrelativistic
 and a
relativistic two-nucleon Schr\"odinger equation is proposed which 
allows to  analytically rewrite  the two Schr\"odinger equations into each 
other. As a consequence a NN potential occurring in the relativistic 
 Schr\"odinger equation  can be gained from a nonrelativistic  one by an 
analytical procedure.
The S-matrices in the two equations are exactly identical and therefore the 
two-nucleon phase shifts. 

\end{abstract}


\narrowtext

%
Few-nucleon equations and their numerical solutions are mostly  carried
out  in  a 
nonrelativistic  framework. Fitting  NN potentials to  NN data
 incorporates to some extent relativistic features into the non-relativistic 
framework, 
  in other words relativistic features contained in the data are absorbed
  in the potential parameters of the nonrelativistic Schr/"odinger equation.
This however is not  sufficient, 
 especially relativistic effects in systems with more than
 two particles are not accessible in this manner. 
As a first step towards a relativistic framework the NN system in its 
 center of mass frame should have the correct form \cite{bak,fong},
 which requires that  
the operator for the kinetic energy is formed out of square roots. 
The present day so called realistic  NN forces are fitted to NN data 
together with the nonrelativistic form of the kinetic energy
\cite{cdbonn,nijm,av18}. 
More precisely a mixed procedure has been applied. These potential models
  were fitted to the phaseshifts and amplitudes of the Nijmegen partial- wave
  analysis \cite{NEWREF1}, 
whereby relativistic relations between the c.m. momenta and
  the kinetic energy and between the differential cross section and the
  scattering amplitude have been used. Clearly  consistent steps would be
  desireable and to refrain altogether from using the nonrelativistic NN 
  Schr\"odinger equation.
 The question addressed in this paper is,  whether the potentials have to
be refitted to the data once the relativistic form of the kinetic energy
is used instead of the nonrelativistic one. 
 A refitting of the NN potential has been undertaken e.g. in \cite{car}. 
In earlier work \cite{glo,coe} approximate relations among the potentials
 in the 
nonrelativistic and relativistic Schr\"odinger equations were introduced.
In Ref\cite{kei} relations among the squares of the interacting and the free
mass operators are used to find formal relations between the
relativistic and nonrelativistic Schr\"odinger equation. However these
formal relations do not allow to write the potential in the relativistic
Schr\"odinger equation explicitely in terms of the potential in the
nonrelativistic Schr\"odinger equation.
 We want to propose here a procedure providing an exact analytic relation
between those potentials.  
  Before, however, we would like to emphasize that we are not generating
  a NN potential which has all the correct relativistic features built in
  ( like an effective potential derived from field theory in the Hamiltonian
  formalism \cite{NEWREF2}) 
but simply establish a formal connection among potentials
  in the relativistic and nonrelativistic forms of the Schr/"odinger equation.
  The potentials are treated as " black boxes" simply as functions of
  certain momenta with no reference to whether the potentials had originally
  some relativistic background or not.

In practice few-nucleon problems are mostly solved in a partial wave
representation. Therefore we formulate that transformation among the
potentials for given angular momentum states.
Let $l$, $s$ and $j$ denote the orbital, total spin and total angular momenta 
of the NN system, then for a given $j$ and $s$ the nonrelativistic
 Schr\"odinger equation in momentum  space reads

\begin{eqnarray}
\left( { p^2 \over m} + 2m \right)  \psi_l (p)  +
\sum _{l'} \int_0 ^\infty dp'  {p'}^2  V_{l l '} (p,p') \psi_{l '} (p
') = E \psi_l (p) 
\label{eq1}
\end{eqnarray}
Here $m$ is the nucleon mass and the sum over $l'$ is present or not depending 
on $s$ and $j$. 
In Eq.~(\ref{eq1}) the rest masses $2m$, which  are usually absorbed into
the definition of the energy, are still contained.
The relativistic version of the two-nucleon Schr\"odinger equation  is
given by  \cite{bak,fong,kei} 
\begin{eqnarray}
2\sqrt{m^2 + p^2} \phi_l (p) + \sum_{l'} \int _0^\infty dp' {p'}^2 
U_{l l'} (p,p') \phi_{l '} (p')  = E \phi_l (p) 
\label{eq2}
\end{eqnarray}
The question is, if one can  find the potential $U$  in Eq.~(\ref{eq2}), once 
the potential $V$ in the nonrelativistic Schr\"odinger 
equation  Eq.~(\ref{eq1}) 
is given. 

We propose the  use of two types of  momenta 
 and  denote the momenta in the nonrelativistic 
Eq.~(\ref{eq1}) by $q$. 
We define the following  relation between the momenta q  in Eq.~(\ref{eq1}) 
and  p in Eq.~(\ref{eq2}):
\begin{equation}
2 \sqrt{ m^2 + p^2 } \equiv 2m + { { q ^2 } \over m } 
\label{eq3}
\end{equation}
This relation can be solved  either for p or for  q :
\begin{equation}
p = q \sqrt{ 1 + { {q^2} \over { 4 m ^2 } } } 
\label{eq4}
\end{equation}
and
\begin{equation}
q = \sqrt{2m} \sqrt{ E_p - m } 
\label{eq5}
\end{equation}
where we  define  $E_p= \sqrt{m^2 + p^2}$.
For small values of  $p$ or $q$ one  obtains 
\begin{equation}
p \approx 
q ( 1 + { { q^2 } \over { 8 m ^2 } } ) 
\label{eq6}
\end{equation}
and
\begin{eqnarray}
q \approx
p (  1 - { { p^2 } \over { 8 m ^2 } } )
\label{eq7}
\end{eqnarray}
whereas for large  values of $p$ and $q$ one gets 
\begin{eqnarray}
p \approx { { q^2 } \over { 2m} } 
\label{eq8}
\end{eqnarray}
 With these definitions it is possible to rewrite Eq.~(\ref{eq1}) into
Eq.~(\ref{eq2})
and vice versa.
Using Eqs.~(\ref{eq3}) and (\ref{eq4}) one  obtains
\begin{eqnarray}
p^2 dp &=& q^2 dq \sqrt{ 1 + { { q^2} \over { 4 m^2 } } } ( 1 + { { q^2} 
\over { 2 m^2 } } ) \cr
&\equiv&  q^2 dq h^2(q)
\label{eq9}
\end{eqnarray}
with
\begin{equation}
h(q) \equiv \sqrt{ \left( 1 + \frac {q^2}{2 m^2} \right) 
\sqrt{ 1 + \frac {q^2}{4 m^2}} } 
\label{eq10}
\end{equation}
Then Eq.~(\ref{eq2}) turns into
\begin{eqnarray}
\left( 2m + { {q^2 } \over { m }}  \right) \phi_l (p) 
+ \sum_{l'} \int _0 ^\infty dq' {q'}^2 h^2 (q') U _{l l'} (p,p') \phi_{l'}
(p') = E \phi_l (p) 
\label{eq11}
\end{eqnarray}
If we define 
\begin{eqnarray}
\psi_l(q) \equiv h(q) \phi_l(p) 
\label{eq12}
\end{eqnarray}
and
\begin{eqnarray}
V_{l l'}(q,q') \equiv h(q) U_{l l'} (p,p') h(q') 
\label{eq13}
\end{eqnarray}
we  arrive at  
\begin{eqnarray}
\left( 2m + { {q^2 } \over m} \right) \psi_l (q) + \sum _{l'}
\int _0 ^\infty dq' {q'}^2 V_{l l'} (q,q') \psi_{l'} (q') 
= E \psi_{l} (q)
\label{eq14}
\end{eqnarray}
which is the nonrelativistic Eq.~(\ref{eq1}). Thus Eqs.~(\ref{eq12}) and 
(\ref{eq13}) provide the  desired relations  between the  wave functions
and  the potentials. 
 Explicitly written , the potential  $U$ results from $V$  via 
\begin{eqnarray}
U_{l l'} (p,p') &=& h^{-1}(q) V_{l l'}(q,q')h^{-1} (q') 
\cr
&=& { { \sqrt{2} m } \over \sqrt{ E_p \sqrt{ 2m^2 + 2m E_p}} } 
V_{l l'} (\sqrt{2m}\sqrt{E_p -m} , \sqrt{2m}\sqrt{E_{p'}-m} ) 
    { { \sqrt{2} m } \over \sqrt{ E_{p'} \sqrt{ 2m^2 + 2m E_{p'}}} } 
\label{eq15}
\end{eqnarray}
and the potential $V$ results from $U$  via 
\begin{eqnarray}
V_{l l'} ( q , q' ) &=& 
\sqrt{ ( 1 + { { q^2 } \over { 2 m^2 } } ) \sqrt{ 1 + {{ q^2 } \over { 4 m^2}}}
}
\cr
&~& U_{l l'} ( \sqrt{ q^2    + { { q^4 } \over { 4 m^2 }  } } , 
               \sqrt{ {q'}^2 + { { {q'}^4 \over { 4 m^2} }} } ) 
\sqrt{ ( 1 + { { {q'}^2 } \over { 2 m^2 } } ) 
\sqrt{ 1 + {{ {q'}^2 } \over { 4 m^2}} } }
\label{eq16}
\end{eqnarray}
Correspondingly, the relativistic wave function expressed in terms of the 
nonrelativistic one is given by
\begin{eqnarray}
\phi_l(p) = { { \sqrt{2} m} \over { \root 4 \of { 2m ^2 + 2m E_p }
\sqrt{ E_p } }  } \psi_l ( \sqrt{ 2m } \sqrt{ E_p -m } ) 
\label{eq17}
\end{eqnarray}
The transformation  of the wave function given in Eq.~(\ref{eq12})
 conserves the norm. We have
\begin{eqnarray}
\int _0 ^\infty dp p^2 \phi_l ^2 (p) = \int _0 ^\infty dq q^2 h^2 (q) 
\phi_l^2 (p)  =  \int _0 ^\infty dq q^2 \psi_l ^2 (q ) 
\label{eq18}
\end{eqnarray} 

Let us now  consider the Lippmann Schwinger equations for the partial wave 
projected half-shell t-matrices. The relativistic version is  given by
\begin{eqnarray}
T_{l l'} (p,p') = U_{l l'} ( p,p') + 
\sum_{l''}\int _0 ^\infty dp'' {p''}^2 U_{l l''} (p,p'') { 1 \over 
{ 2E_{p '} - 2E_{p''} + i \epsilon } } T_{l'' l'} (p'',p') 
\label{eq19}
\end{eqnarray}
Using Eqs.~(\ref{eq9}) and (\ref{eq13}) and defining 
\begin{eqnarray} 
t_{l l'} (q ,q') = h(q) T_{l l'} (p,p') h(q') 
\label{eq20}
\end{eqnarray}
we  obtain  from  Eq.~(\ref{eq19}) 
\begin{eqnarray}
t_{l l'} (q,q') = V_{l l'} (q, q') + \sum _{l''} \int _{0 } ^{\infty} 
dq'' {q''} ^{2 } V_{l l''} (q,q'') { 1 \over { { {q{ '}}^{2} \over m} - 
{{q''}^{2} \over m } + i\epsilon  }  } t_{l'' l'} (q'' , q' ) 
\label{eq21}
\end{eqnarray}
 This is the standard nonrelativistic Lippmann Schwinger equation.

 Eq.~(\ref{eq20}) provides also a relation  between the phase shifts
evaluated through the relativistic and nonrelativistic Lippmann Schwinger
equations. The unitarity relation resulting from  Eqs.~(\ref{eq19}) 
and (\ref{eq21})
are  given by 
\begin{eqnarray}
Im T_{l l} (p, p ) = 
- { {\pi p E_{p} } \over 2 } \sum _{l'} \vert T_{l l'} (p
, p) \vert ^{2 } 
\end{eqnarray}
and 
\begin{eqnarray}
Im t_{l l} (q , q) = 
-{ { \pi q m } \over 2 } \sum _{ l'} \vert t _{l l'}
(q , q) \vert ^{2 } 
\end{eqnarray}
As a consequence the S-matrices  given by
\begin{eqnarray}
S_{l l'} ( p)  \equiv \delta_{l l'} - i\pi p E_{p} T _{l l'} (p, p)
\end{eqnarray}
and 
\begin{eqnarray}
s_{l l'}(q)  \equiv \delta_{l l'} - i \pi q m t_{l l'} (q,q) 
\end{eqnarray}
are unitary. 
Using Eq.~(\ref{eq20}) and the relations (\ref{eq3}) and
(\ref{eq4}) one  obtains 
\begin{eqnarray} 
s _{l l'} (q) 
&=& \delta_{l l'} - i \pi q m h(q) T_{l l'} (p, p) h(q)
\cr
&=& \delta_{l l'} - i \pi q m \sqrt{ 1 + {{ q^{2} } \over {4 m^{2}} } } 
( 1 + {{q^{2}} \over {2m^{2}}})    T _{l l'} (p ,p) 
\cr
&=&  \delta_{l l'} - i \pi 
q m { p \over q } { 1 \over {2m}  } ( 2m +{ {q^{2}}
  \over m} ) T _{l l'} (p, p) 
\cr
&=&\delta_{l l'} - i \pi p E_{p} T_{l l'} (p , p) =S_{l l'} (p)
\label{eq26}
\end{eqnarray}
 From Eq.~(\ref{eq26}) we see that the S-matrices are identical, and as
 a consequence  the scattering
phase shifts and mixing parameters are identical. Though the momenta q
and p for the nonrelativistic and relativistic S-matrices are different,
the related energies are the same, namely
$E=2 m + q^2/m = 2 \sqrt{m^2 + p^2}$.

In Fig.~1 we show the  relation  between the momenta $p$ and
$q$ as given in Eq.~(\ref{eq4}). In Fig.~2 we display  
the s- and d-wave components of the deuteron wave function. 
We  show  these wave function components as a function of their respective
momenta, $p$ in the relativistic case and $q$ in the nonrelativistic case.
The underlying NN potential was arbitrarily chosen as the  CD Bonn  potential
\cite{cdbonn}. The minima for the s-and d-wave components are shifted
towards larger momenta in the relativistic case. The effects are large
above about 5 $fm^{-1}$.

Summarizing, we have shown that NN potentials fitted to NN data in a
nonrelativistic framework can be analytically rewritten by a scale
transformation in the momenta such that they lead to the same S-matrix when
 used in a relativistic Schr\"odinger equation. There is no need for
refitting parameters when the nonrelativistic operator of kinetic energy is
replaced by the relativistic square root expressions. 

 The potential $U$ can be used in few-nucleon systems with $ A > 2 $ as
done
 in the studies \cite{car,glo}.

 Acknowledgment: This work was supported by  the Research Contract
No. 41324878 (COSY-044) with the Forschungszentrum J\"ulich, Germany.
The authors would like to thank Ch.~Elster for critically reading the
manuscript.


\begin{figure}
\centering
\mbox{\epsfysize=100mm \epsffile{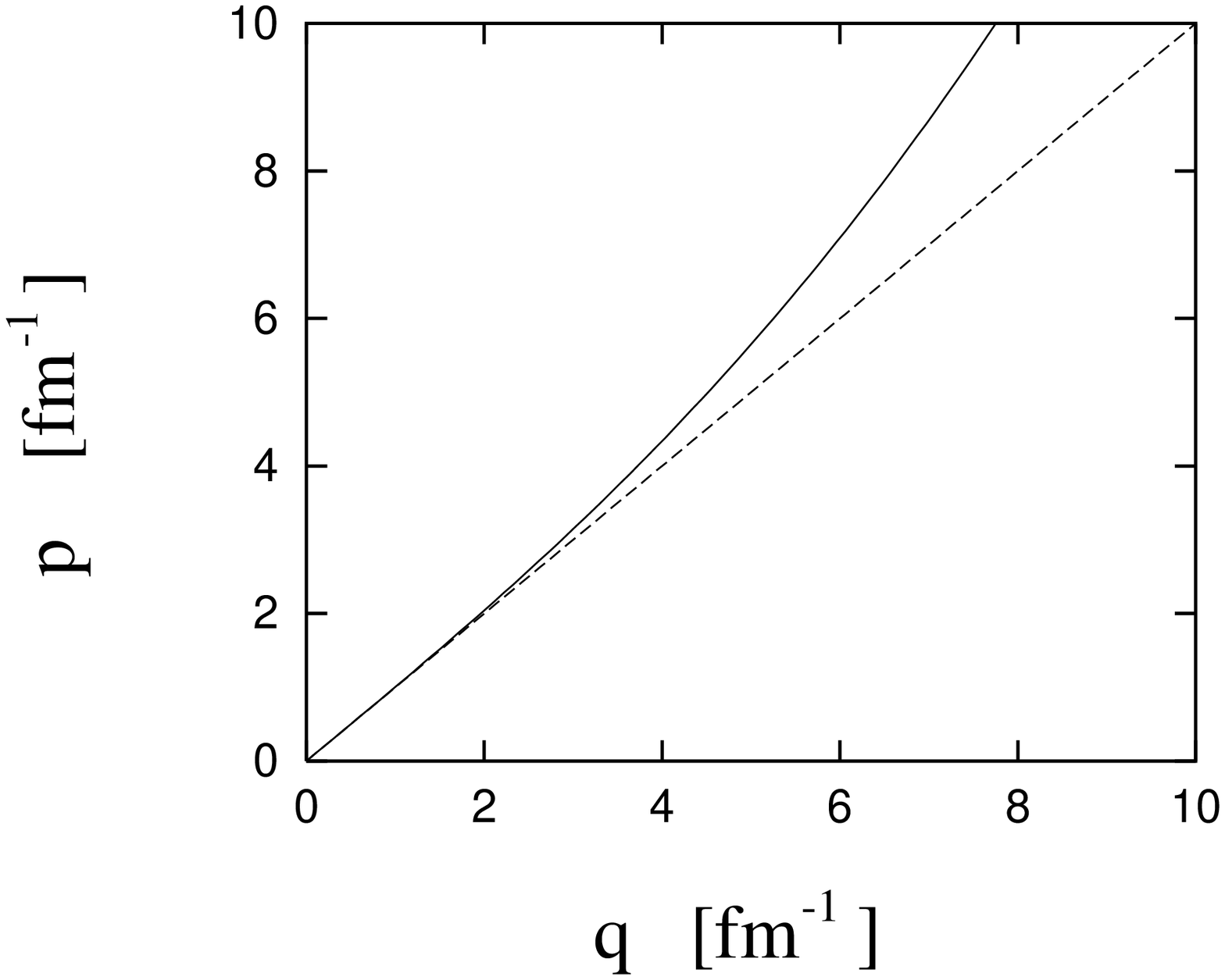}}
\caption{ The relation between the 
relativistic momentum $p$ and the nonrelativistic momentum  $q$ 
as given in Eq.~(4). 
}
\label{f1}
\end{figure}

\begin{figure}
\centering
\mbox{\epsfysize=100mm \epsffile{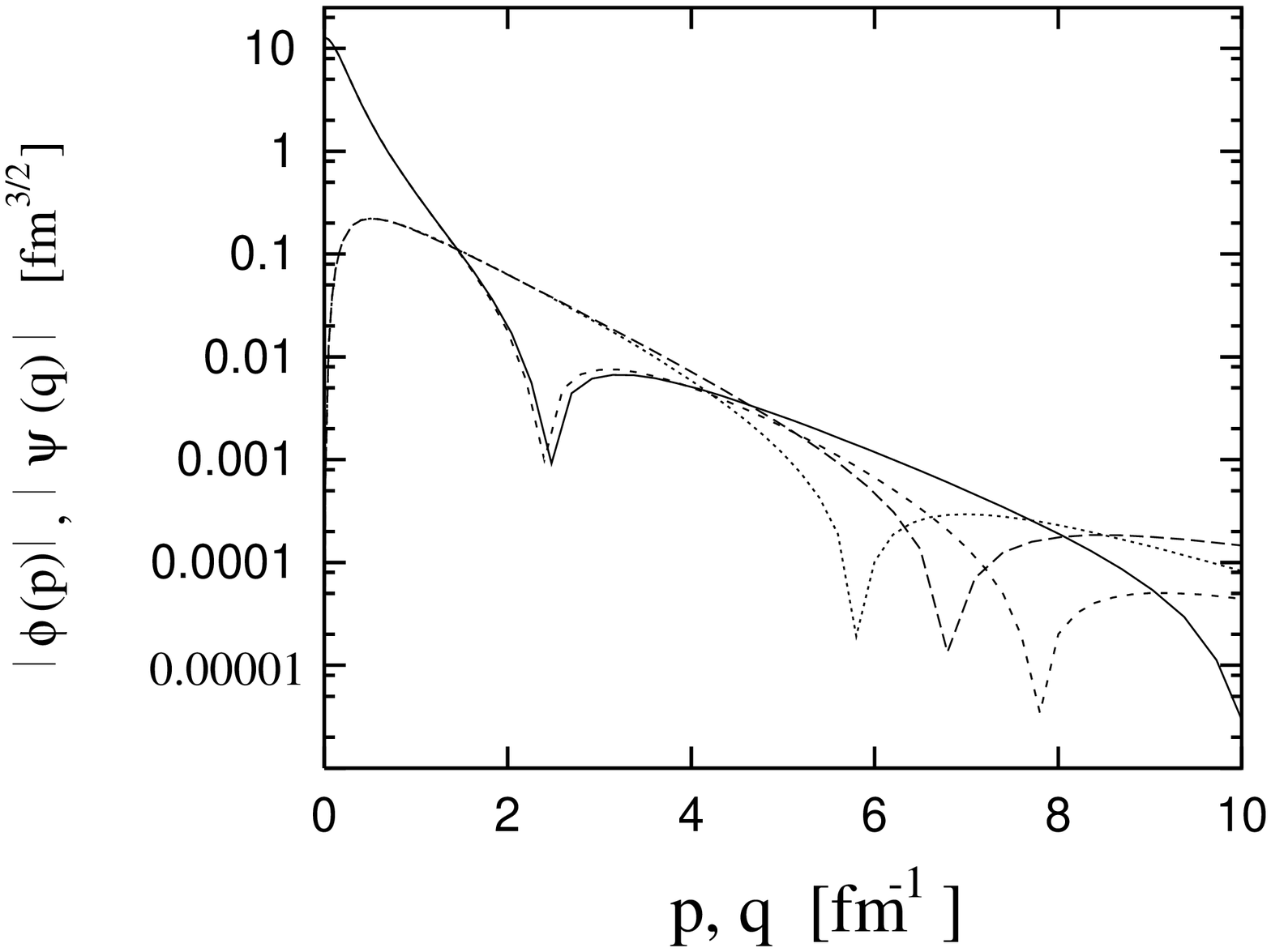}}
\caption{ The absolute values of the  relativistic and 
nonrelativistic wave functions,
$\phi (p)$ and 
$\psi (q) $ for the deuteron bound state. 
The s-wave components  $\phi_0 (p)$ (solid curve) and $\psi_0 (q)$ 
(short dashed curve) and the d-wave components
 $\phi_2 (p)$ (long dashed curve)  and $\psi_2 (q)$ (dotted curve) 
have large differences at higher momenta.
}
\label{f2}
\end{figure}

\end{document}